\documentclass[jkps,preprint,fleqn,showpacs,showkeys]{revtex4}
\usepackage[pdftex]{graphicx}
\usepackage{amssymb}
\usepackage{amsmath}
\usepackage{bm}

\begin{document}
\setcounter{page}{1}
%
\title{Quark-Gluon Jet Discrimination Using Convolutional Neural Networks}
\author{Jason Sang Hun \surname{Lee}}
\email{jason.lee@uos.ac.kr}
\affiliation{Department of Physics, University of Seoul, Seoul 02504}
\author{Inkyu \surname{Park}}
\affiliation{Department of Physics, University of Seoul, Seoul 02504}
\author{Ian James \surname{Watson}}
\email{ian.james.watson@cern.ch}
\affiliation{Department of Physics, University of Seoul, Seoul 02504}
\author{Seungjin \surname{Yang}}
\email{seungjin.yang@cern.ch}
\affiliation{Department of Physics, University of Seoul, Seoul 02504}
\thanks{Fax: +82-2-554-1643}

\begin{abstract}
Currently, newly developed artificial intelligence techniques, in
particular convolutional neural networks, are being investigated for
use in data-processing and classification of particle physics collider
data. One such challenging task is to distinguish quark-initiated jets
from gluon-initiated jets. Following previous work, we treat the jet
as an image by pixelizing track information and calorimeter deposits
as reconstructed by the detector. We test the deep learning paradigm
by training several recently developed, state-of-the-art convolutional
neural networks on the quark-gluon discrimination task. We compare the
results obtained using various network architectures trained for quark-gluon
discrimination and also a boosted decision tree (BDT) trained on summary
variables.
\end{abstract}
  
\maketitle

\section{Introduction}

In recent years, new techniques have been developed for image analysis and classification. 
This arose from the finding that neural networks composed of successive layers of convolutional filters operating on the previous layer can be successfully and efficiently trained with networks dozens of layers deep.
This is to be compared to traditional densely connected neural networks, where adding multiple layers makes training unstable and slow and generally does not improve performance.
This new paradigm of a neural network goes under the moniker of Deep Learning and was most famously adapted for use in the creation of AlphaGo, a Go AI which achieved the unprecedented feat of defeating a Go world champion \cite{silver2016mastering}.


Recently, these techniques have been applied to jet physics by interpreting the energy depositions forming a two-dimensional image in $(\eta, \phi)$ space~\cite{Cogan:2014oua, de2016jet}. The space is pixelized, and the pixel luminosity is proportional to the amount of energy carried by particles of the jet travelling in the direction of the pixel.
Convolutional neural network techniques were applied to these jet-images~\cite{Cogan:2014oua}.
This has been extended by treating the different types of particles as being different color channels producing a color image representation of the jets \cite{Komiske:2016rsd}.
In this paper, we train various recently-developed, start-of-the-art convolutional neural network types to discriminate quark-initiated jets and gluon-initiated jets and compare the results from the different networks. In particular, we discuss the expected performance from 13 TeV LHC data with the CMS detector \cite{Chatrchyan:2008aa}.
Other approaches being investigated for jet physics include geometric deep learning for processing non-Euclidean data, such as graphs and manifolds \cite{bronstein2017geometric}. The authors of these studies argue that this approach can reduce the loss of information that occurs with pixelization so that classification performance can be improved \cite{louppe2017qcd,cheng2018recursive}.

\section{Monte Carlo Models}

We use MadGraph5\_aMCatNLO v2.6.0 to generate the hard process for dijet and Z+jet events at leading order \cite{Alwall:2014hca}. We separately generate events for quarks and gluons and label the jets ``quark'' or ``gluon'' based on the hard process being generated. We do this to avoid ambiguities in the matching process. As outlined in the event selection below, we also require dijets to be balanced to reduce further ambiguities due to hard radiation producing further gluon-like jets.
We interface the generated hard-process events to PYTHIA 8.2 with the default PYTHIA tune for parton showering and underlying event generation \cite{Sjostrand:2014zea}.
We use the fast detector simulator DELPHES to approximate CMS reconstruction particle-flow algorithms \cite{deFavereau:2013fsa}. DELPHES uses the FASTJET package for anti-$\it{k_{T}}$ algorithm with a jet radius of R = 0.5 \cite{Cacciari:2011ma}.
The default settings of the packages have been used to generate events.

\section{Event Selection}

We apply event requirements which follow the event selection for the CMS 13 TeV quark-gluon BDT discrimination \cite{CMS-PAS-JME-16-003}.
For dijet events, we require that the two jets be \emph{balanced}. At least two jets in the event are required to have transverse momentum \(p_T\) greater than 30 GeV.
The azimuthal angle between the two leading jets \(\Delta\phi\) is required to satisfy \(\Delta\phi > 2.5\). 
If a third jet is reconstructed in the event, it must have \(p_T\) less than 30\% of the average \(p_T\) of the two leading jets.
We study both of the two leading jets from events passing this selection.

Similarly, for Z+jet events, the jet must balance against the Z boson without additional high-energy activity. Two muons are required to be reconstructed with the dimuon invariant mass within 20 GeV of the Z boson mass and at least one jet of 30 GeV is required. 
The azimuthal angle \(\Delta\phi\) between the Z and the leading jet must satisfy \(\Delta\phi > 2.5\). 
Any additional jets that are reconstructed must have less the 30\% of the \(p_T\) of the dimuon system.
We study only the jet balanced against Z boson for these event.


The jet production cross-section falls logarithmically with increasing hard parton \(p_T\).
Samples were, therefore, generated requiring that the \(p_T\) of the generated hard partons satisfy minimum and maximum conditions in 100 GeV increments in order to obtain representative samples for all \(p_T\) ranges we would expect to observe at the LHC. 

%

\section{Event Processing}

After the events have been selected, the jets are processed to a form suitable representation for the inputs to the convolutional neural networks.
A $33\times 33$ \((\eta, \phi)\) grid is constructed representing a total size of \((\Delta\phi, \Delta\eta) = (0.8, 0.8)\), corresponding to a jet cone of \(\Delta R = 0.5\), where the central bin is aligned to the jet axis. For each particle within the jet, the particle $p_{T}$ is added to the bin corresponding to the particle momentum in \((\Delta\eta, \Delta\phi)\) relative to the jet axis.

\begin{figure}
\includegraphics[width=1.0\textwidth]{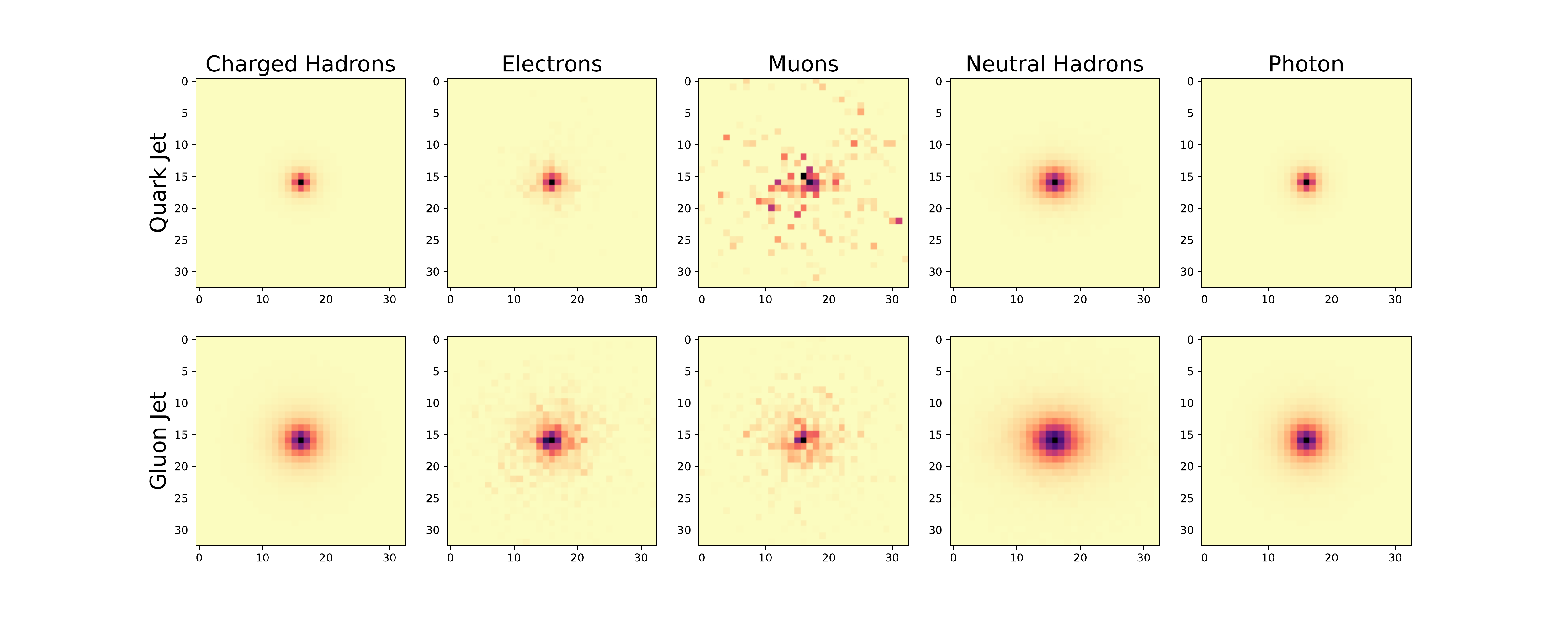}
\caption{Summed jet images divided into the various channels used in the analysis for the \(p_t\) range from 100 to 200 GeV.}
\label{fig:jet-images}
\end{figure}

Several such channels have been filled corresponding to different particle types. In the standard case, we have separate channels for charged particle \(p_T\), neutral particle \(p_T\), and number of charged particles (i.e., the pixel luminosity represents simply the number of charged particles in the grid bin, rather than the summed \(p_T\)). We have also tested adding additional channels for more granular information. We split the charged particles into electron, muon, and charged hadron categories, and the neutral particles into photon and neutral hadron categories.
These represent all the separable particle types available from the CMS detector output.
For each of these categories, we create channels from both the number of particles and the particle \(p_T\) for a total of ten channels.

We have also tested the effects of varying the size of the grid and the nonlinear scaling of the bin widths.
After filling each pixel grid, we normalize each channel to the maximum value for a given sample.
Figure~\ref{fig:jet-images} shows summed jet images from the various channels used in the analysis.

\section{Neural Network Models}

We implemented several state-of-the-art neural networks in the Keras framework \cite{chollet2015keras}. In this section, we briefly describe the characteristics of these networks.

VGGNet uses only small $3\times3$ filters and has many hidden layers compared with other ConvNet models. This architecture increases the nonlinearity of the model \cite{DBLP:journals/corr/SimonyanZ14a}.
The deep residual network (ResNet) uses identity mappings as skip connections, which enables information flows across layers \cite{DBLP:journals/corr/HeZRS15, DBLP:journals/corr/HeZR016}.
The inception network (GoogLeNet) and its update (inception-v4) including ResNet components (inception-resnet) are stacked inception modules which has the effect of making the neural network wider and deeper with fewer parameters and lower computational cost through dimensionality reduction and the usage of various sized filters \cite{DBLP:journals/corr/SzegedyLJSRAEVR14, DBLP:journals/corr/SzegedyIV16}.
The densely connected convolutional network (DenseNet) consists of dense blocks, where each layer is directly connected to all preceding layers \cite{DBLP:journals/corr/HuangLW16a}. 
Xception consists of stacked depthwise separable convolution layers. A depthwise separable convolution layer consists of a depthwise convolution followed by a pointwise convolution without intermediate activation. Eliminating intermediate non-linearity results has been experimentally seen to improve the model's learning speed and performance \cite{carreira1998xception}.
The Squeeze-and-Excitation Network (SENet) is designed to analyze the relationship between channels of the feature maps. SENet is implemented by applying an SE block (which extracts information per channel from a feature map and then recalibrates it) to other architectures. In this paper, we use the Inception and ResNet networks (inception-resnet) for the implementation of SENet \cite{DBLP:journals/corr/abs-1709-01507}.

We have also implemented a simple convolutional network Vanilla ConvNet, which contains a convolution filter layer and a ReLU activation layer, then three blocks consisting of a batch normalization layer, then an activation layer, and then a convolutional filter layer. These are then passed to a Global Average Pooling layer and finally a softmax output. This Vanilla convnet is more closely modelled on previously studied networks for comparison with the state-of-the-art networks described above.

\section{Results}

\begin{figure}
\includegraphics[width=0.69\textwidth]{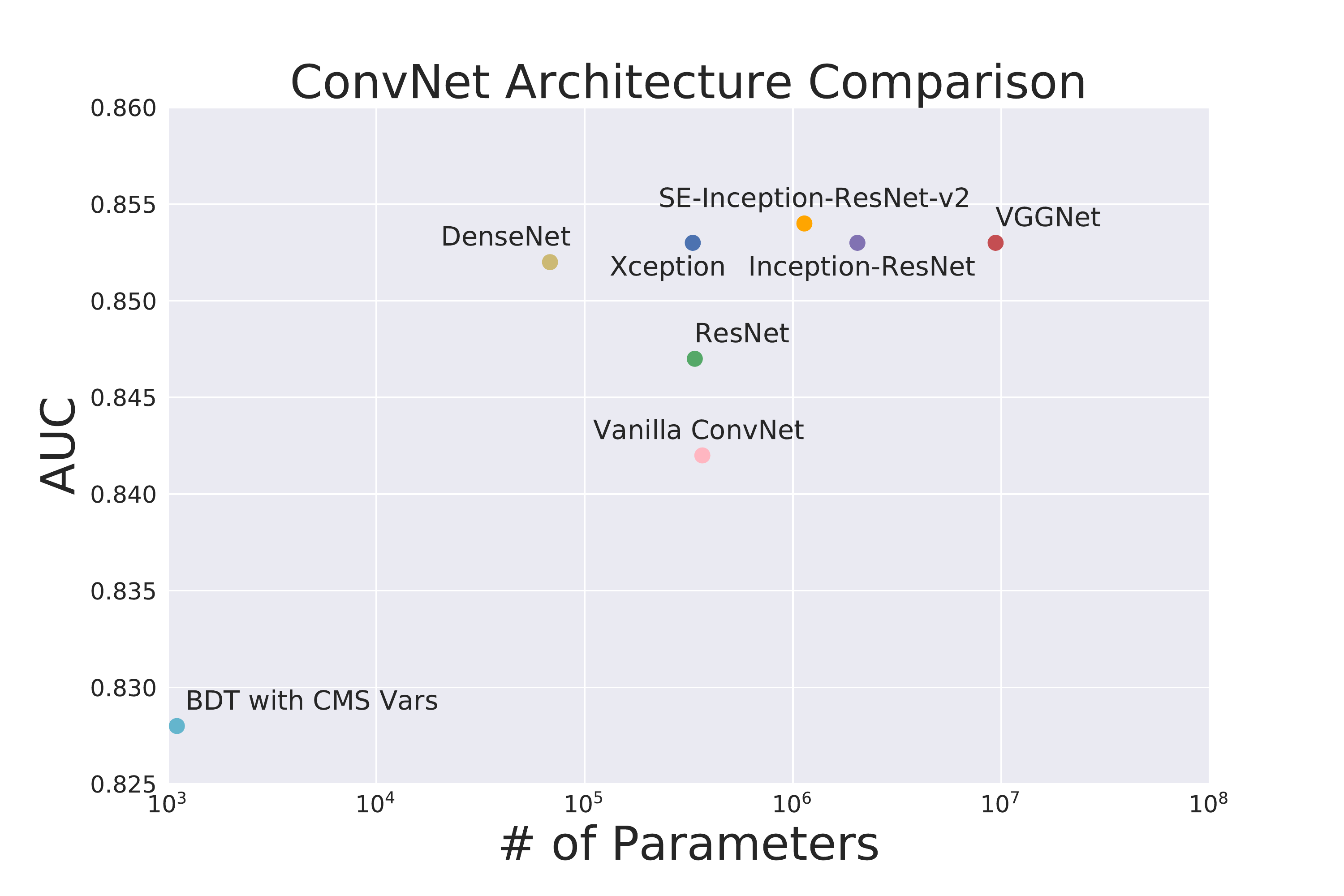}

\caption{Best obtained area under curve (AUC) versus number of parameters for various state-of-the-art convolutional neural networks trained on jets with a \(p_T\) range from 100 to 200 GeV and split into 10 channels.}
\label{fig:sota_covnet}
\end{figure}

The various convolutional networks were trained on 10-channel samples with \(p_T\) in from 100 to 200 GeV, and the resulting test AUC distributions are shown in Figure~\ref{fig:sota_covnet}. 
The figure shows that there is a saturation point is reached by even networks with low numbers of parameters. Therefore, we restrict subsequent explorations to a simple convnet.

\begin{figure}
\includegraphics[width=0.69\textwidth]{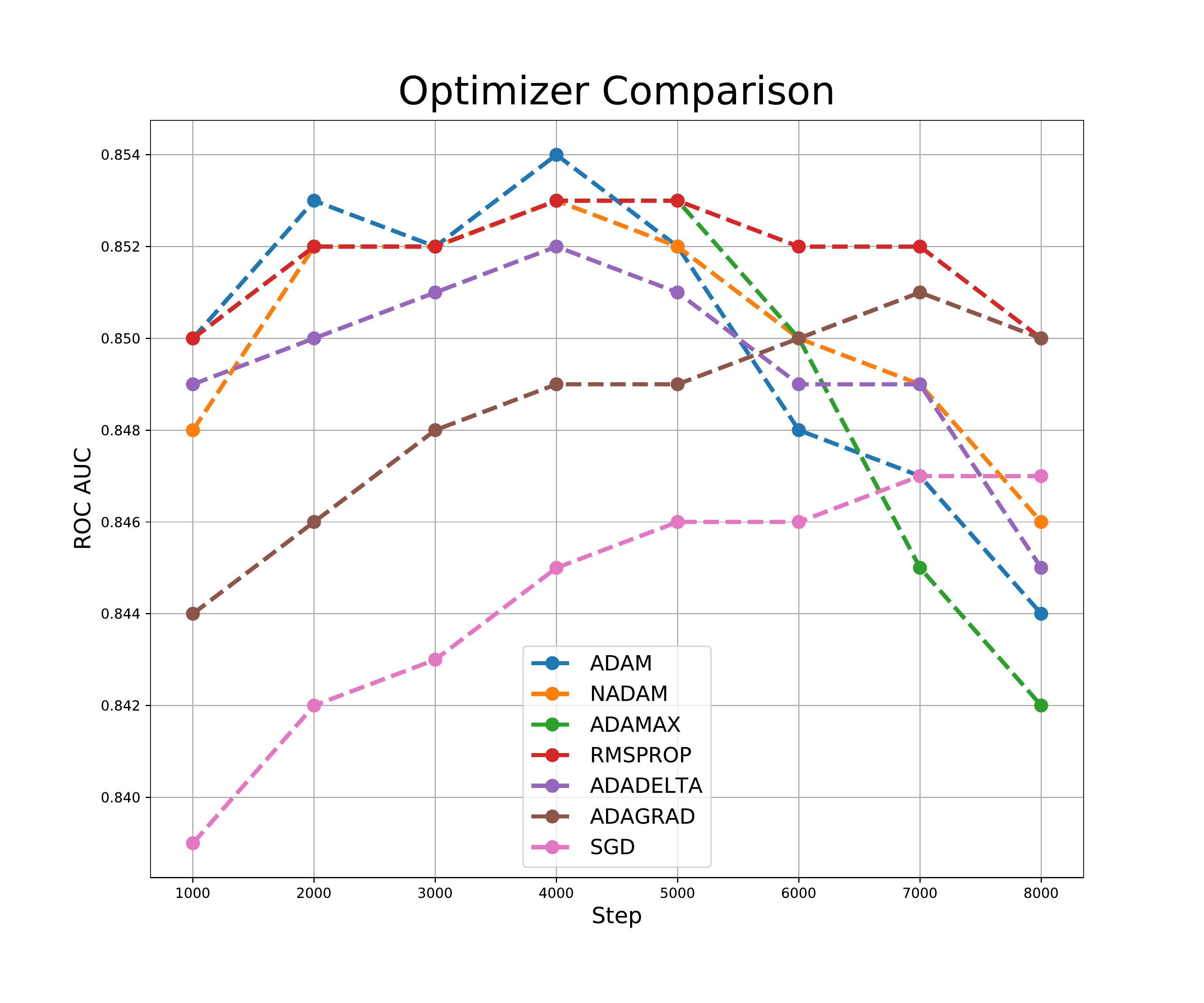}

\caption{Area under curve (AUC) versus number of batch steps while training using different optimizers.}
\label{fig:optim_com}
\end{figure}

Figure~\ref{fig:optim_com} shows the area under curve (AUC) versus the number of batch steps for training using various optimizers.
Training was done on the 4-block Vanilla ConvNet network with 8 image channels, each of size $33\times 33$, and with a batch size of 512. 
Default arguments are used. RMSProp is seen to be the most stable. 

\begin{figure}
\includegraphics[width=0.69\textwidth]{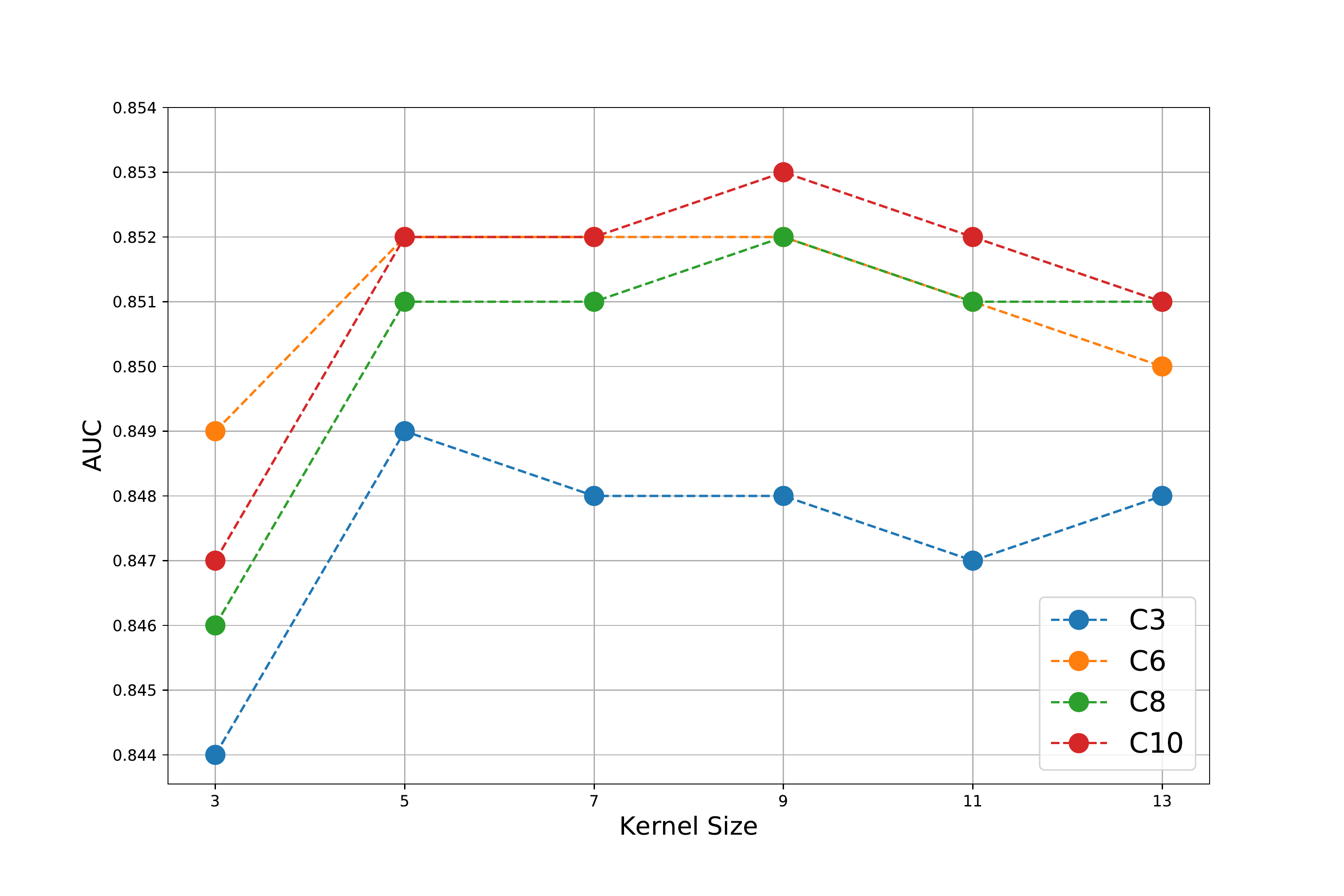}

\caption{Best area under curve (AUC) versus number of channels and kernel size, which is obtained by using the Vanilla ConvNet setup with jets in the \(p_T\) range from 100 to 200 GeV.}
\label{fig:rgb-channels}
\end{figure}

Using the same basic ConvNet setup, we also explored the input data by increasing the number of data channels.
The channels represent the sum \(p_T\) or the number of particles for various categories of particles observable by CMS.
These categories are: charged hadrons, neutral hadrons, electrons, photons and muons.
Further combinations of these are possible to reduce the number of channels.
The electrons and muons can be combined into a lepton channel, the neutral hadron and photons into a neutral channel, and the charged hadrons and leptons into a charged channel. The networks studied are identified by number of channels. C3 uses charged particle \(p_T\), neutral particle \(p_T\) and charged particle multiplicity. 
Network C6 uses charged particles, neutral hadrons, and photons, each with channels for \(p_T\) and multiplicity. C8 breaks up the charged particles so has channels for charged hadrons, leptons, neutral hadrons, photons in both \(p_T\) and multiplicity. Finally, C10 has channels for charged hadrons, electrons, muons, neutral hadrons, and photons in \(p_T\) and multiplicity.
Figure~\ref{fig:rgb-channels} shows the resulting best AUC using networks with different numbers of channels, and also varying the kernel size of the convolutional filter. Increasing the filter size from 3x3 to 5x5 results in increased performance for all channels, while further increases of size show no stable gain or loss. Increasing from the 3-channel to the 6-channel setup also provides improvement in classification ability, and appears to saturate the useful information available to the network.

\begin{figure}[t]
\includegraphics[width=0.69\textwidth]{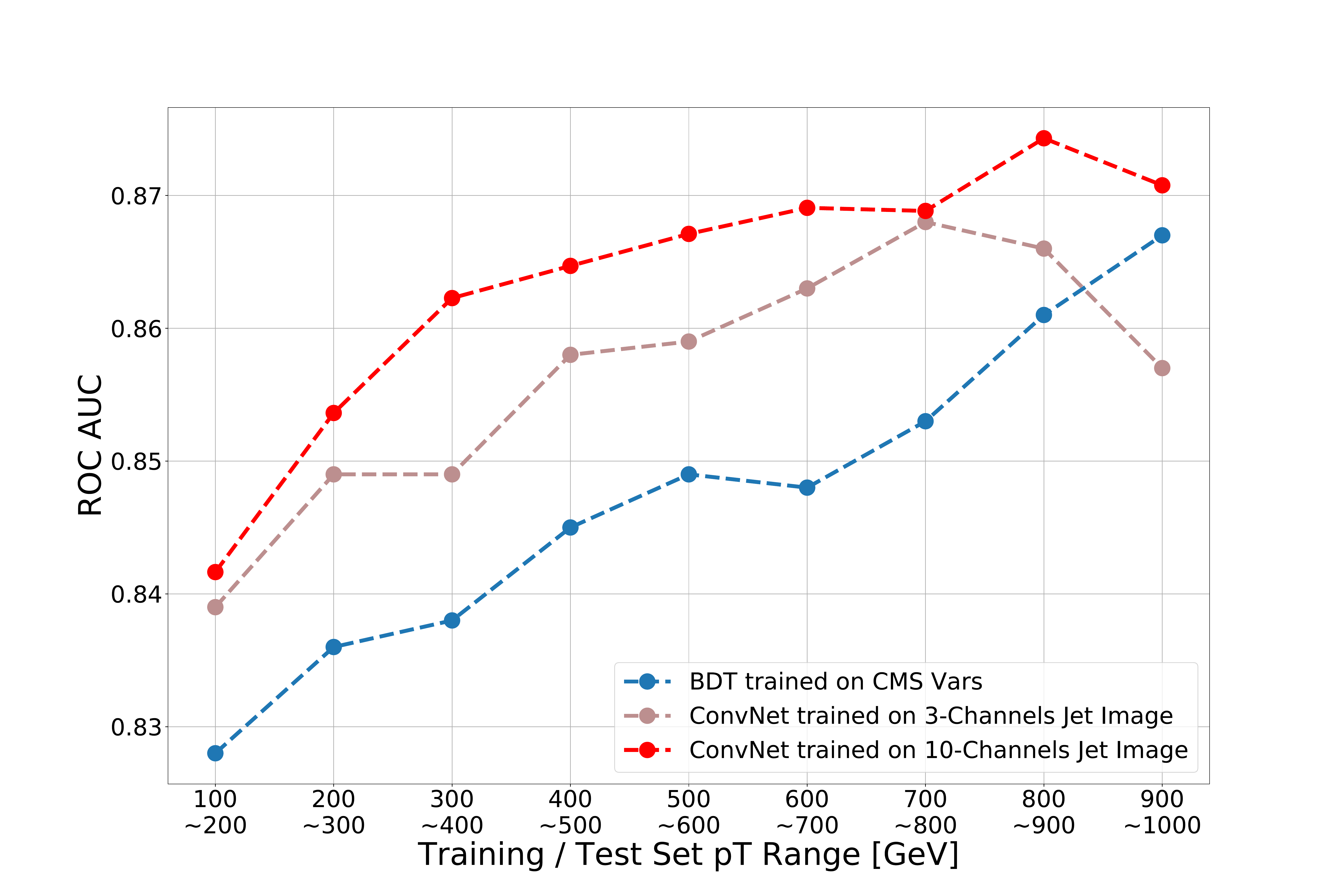}

\caption{Best obtained area under curve (AUC) versus number of channels and \(p_T\). Also included is the AUC for a BDT trained on several variables and used by the CMS collaboration.}
\label{fig:pt-channels}
\end{figure}

The channel variation is further explored in Figure~\ref{fig:pt-channels}, which shows the variation in performance for the networks (using kernel size 5x5) versus the input \(p_T\) range of the jets studied.
Also compared is the BDT, made from five variables, that is typically used by CMS.
The datasets each have about one hundred thousand events.
the ConvNet networks can be seen to have superior performance over the entire \(p_T\) range and to perform particularly well at low \(p_T\).
The 10-channel network also performs better than the 3-channel network over the entire range.

\begin{table}[tbp]
	\begin{tabular}{| c | c | c | c | c |}
\hline
Jet Level   & Model   & {$\epsilon_{g} (\%) \ at \ \epsilon_{q} = 50 \% $}  & {$\epsilon_{g} (\%) \ at \ \epsilon_{q} = 70 \% $} \\
\hline
Particle	& BDT of all jet varaibles \cite{Komiske:2016rsd}      & 5.2 & - \\
            & Shallow Dense Network \cite{Komiske:2016rsd}         & 5.5 & - \\
            & Deep CNN without Color \cite{Komiske:2016rsd}        & 4.8 & - \\
            & Deep CNN with Color \cite{Komiske:2016rsd}           & 4.6 & - \\
\hline
Detector    & BDT with CMS variables                  			   & 6.4 & 17.9 \\
            & VanillaConvNet with 3 channels          			   & 5.8 & 15.6 \\
            & VanillaConvNet with 10 channels       			   & 5.7 & 14.5 \\
            & DenseNet with 10 channels              			   & 5.2 & 14.0 \\
            & SE-Inception-ResNet-v2 with 10 channels 			   & 4.9 & 13.2 \\
\hline

\end{tabular}
\caption{Gluon jet efficiencies at 50\% and 70\% quark jet efficiencies on the test set for a 200 GeV jet. The results from our studies are presented at the detector level and are compared to the results from previous studies performed at the particle level \cite{Komiske:2016rsd}.}
\label{tab:gluon-eff}
\end{table}


Table~\ref{tab:gluon-eff}, shows our results obtained from detector level events produced with DELPHES as previously described. For comparison, we also show the efficiency previously reported at particle level \cite{Komiske:2016rsd}. Under the state-of-the-art networks with 10-channel setting, we achieved a performance comparable to the results obtained at the particle-level, but which only uses up to 3 channels.

\section{Conclusion}

We have explored the differences between several state-of-the-art deep learning convolutional neural networks for the task of distinguishing quark and gluon hadron jets by using a jet imaging technique.
We found that the information available is already saturated by a modestly-sized DenseNet and that larger-scaled, higher-structured neural networks do not improve the classification performance for the typically sparse jet images.
Nonetheless, a small increase in performance is available compared with a typical BDT classifier as has been previously used in CMS.
We also found that modest increases in training stability can be had by using the RMSProp optimizer.
Finally, further improvements in performance are available by increasing the image channels to finer subdivisions than havw previously been studied, with the caveat that the results from the fast simulation particle flow reconstruction from Delphes will need to be validated with detailed simulation studies.

\begin{acknowledgments}
This work was supported by the Korea Research Fellowship Program through the National Research Foundation of Korea (NRF) funded by the Ministry of Science and ICT (KRF project grant number 2017H1D3A1A01052807). This article was supported by the computing resources of the Global Science experimental Data Hub Center (GSDC) in the Korea Institute of Science and Technology Information (KISTI).
\end{acknowledgments}

\newpage


\begin{references}

\bibitem{silver2016mastering} D.~Silver {\it et al.}, Nature {\bf 529}, 484 (2016).

\bibitem{Cogan:2014oua} J.~Cogan, JHEP {\bf 02}, 118 (2015).

\bibitem{de2016jet} L. de Oliveira. et al. JHEP {\bf 2016}, 069 (2016).

\bibitem{Komiske:2016rsd}  P.~T.~Komiske,  E.~M.~Metodiev, and  M.~D.~Schwartz, JHEP {\bf 1701}, 110 (2017).

\bibitem{Chatrchyan:2008aa} S.~Chatrchyan {\it et al.} [CMS Collaboration], JINST {\bf 083}, S08004 (2008).

\bibitem{bronstein2017geometric} M.~Bronstein {\it et al.}, IEEE {\bf 34}, 18 (2017)

\bibitem{louppe2017qcd} G.~Louppe {\it et al.}, arXiv:1702.00748 (2017).

\bibitem{cheng2018recursive} T.~Cheng, Computing and Software for Big Science {\bf 2}, 3 (2018).

\bibitem{Alwall:2014hca} J.~Alwall {\it et al.}, JHEP {\bf 1407}, 079 (2014).

\bibitem{Sjostrand:2014zea} T.~Sj\"{o}strand {\it et al.}, Comput. Phys. Commun. {\bf 191}, 159 (2015).

\bibitem{deFavereau:2013fsa} J.~de~Favereau {\it et al.}, JHEP {\bf 1402}, 057 (2014).

\bibitem{Cacciari:2011ma} M.~Cacciari, G.~P.~Salam, and G.~Soyez, Eur. Phys. J. C{\bf 72}, 1896 (2012).

\bibitem{CMS-PAS-JME-16-003} CMS Collaboration, CMS-PAS-JME-16-003 (2017).

\bibitem{chollet2015keras} F.~Chollet {\it et al.}, \url{https://keras.io} (2015) [site accessed June 19, 2018].

\bibitem{DBLP:journals/corr/SimonyanZ14a} K.~Simonyan and A.~Zisserman, arXiv:1409.1556 (2014).

\bibitem{DBLP:journals/corr/HeZRS15} K.~He, X.~Zhang, S.~Ren and J.~Sun, arXiv:1512.03385 (2015).

\bibitem{DBLP:journals/corr/HeZR016} K.~He, X.~Zhang, S.~Ren and J.~Sun, arXiv:1603.05027 (2016).

\bibitem{DBLP:journals/corr/SzegedyIV16} C.~Szegedy, S.~Ioffe and V.~Vanhoucke, arXiv:1602.07261 (2016).

\bibitem{DBLP:journals/corr/SzegedyLJSRAEVR14} C.~Szegedy {\it et al.}, arXiv:1409.4842 
(2014).

\bibitem{DBLP:journals/corr/HuangLW16a} G.~Huang, Z.~Liu and K.~Q.~Weinberger, arXiv:1608.06993 (2016).

\bibitem{carreira1998xception} J.~Carreira, H.~Madeira, and J.~G.~Silva, IEEE Trans. Softw. Eng. {\bf 24}, 125 (1998).

\bibitem{DBLP:journals/corr/abs-1709-01507} J.~Hu, L.~Shen and G.~Sun, arXiv:1709.01507 (2017).



\end{references}
\end{document}